# A GPU accelerated mixed-precision Smoothed Particle Hydrodynamics framework with cell-based relative coordinates

*Zirui Mao[1], Xinyi Li[1,2], Shenyang Hu[1], Ganesh Gopalakrishnan[2], Ang Li[1*]*

*Corresponding: Ang.li@pnnl.gov, zirui.mao@pnnl.gov*

[1]Pacific Northwest National Laboratory, Richland, WA, USA
[2]School of Computing, University of Utah, Salt Lake City, UT, USA

**Abstract:** Smoothed Particle Hydrodynamics (SPH) is essential for modeling complex large-deformation problems across various applications, requiring significant computational power. A major portion of SPH computation time is dedicated to the Nearest Neighboring Particle Search (NNPS) process. While advanced NNPS algorithms have been developed to enhance SPH efficiency, the potential efficiency gains from modern computation hardware remain underexplored. This study investigates the impact of GPU parallel architecture, low-precision computing on GPUs, and GPU memory management on NNPS efficiency. Our approach employs a GPU-accelerated mixed-precision SPH framework, utilizing low-precision float-point 16 (FP16) for NNPS while maintaining high precision for other components. To ensure FP16 accuracy in NNPS, we introduce a Relative Coordinated-based Link List (RCLL) algorithm, storing FP16 relative coordinates of particles within background cells. Our testing results show three significant speedup rounds for CPU-based NNPS algorithms. The first comes from parallel GPU computations, with up to a 1000x efficiency gain. The second is achieved through low-precision GPU computing, where the proposed FP16-based RCLL algorithm offers a 1.5x efficiency improvement over the FP64-based approach on GPUs. By optimizing GPU memory bandwidth utilization, the efficiency of the FP16 RCLL algorithm can be further boosted by 2.7x, as demonstrated in an example with 1 million particles. Our code is released at https://github.com/pnnl/lpNNPS4SPH.

# Introduction

As a 'meshfree' computational method, the Smoothed Particle Hydrodynamics (SPH) [1, 2] has gained significant prominence thanks to its remarkable adaptability in handling large deformation problems. This makes it an indispensable tool across diverse applications where large deformations are pivotal, such as battery material processing [3, 4], solid phase processing [5-7], earthquake-induced landslides [5-7], and hydrodynamics in ocean and coastal engineering [8, 9] *etc.*, where large deformations are pivotal. However, the 'meshfree' nature of SPH necessitates frequent nearest neighboring particles searching (NNPS) for capturing particles' interaction. NNPS can typically more than 50% of the computational time, and even surpasses 80% in cases with computationally inexpensive constitutive model. As a result, SPH can be computationally demanding due to the extensive NNPS process, constraining its application to complex high-resolution problems.

To address this challenge, various efficient NNPS algorithms have been developed to alleviate the computational complexity of identifying neighbors for SPH particles. The fundamental approach involves computing distances between a particle and all others within the simulation domain, known as all-list algorithms [10, 11]. While straightforward, this approach becomes infeasible for simulations with a large number of particles due to its high computational cost. The cell link list algorithm [12-15] refines this by partitioning the domain into cells and assigning particles to corresponding cells. As particles move, the particles list in each cell is updated, and when searching for neighbors, the algorithm only considers particles within nearby cells, significantly enhancing efficiency. The tree-based algorithms [16, 17] use hierarchical data structures like octrees or binary trees to partition the simulation domain into regions. Each node of the tree represents a spatial region, and particles are distributed among the nodes. When searching for neighbors, the algorithm recursively traverses the tree to identify relevant nodes and particles, thus narrowing down the search space. Tree-based methods are particularly efficient for simulations with non-uniform particle distributions. Similar to the binary tree approach, the k-d tree (k-dimensional tree) [18, 19] is very efficient in higher dimensions and can adapt to dynamic simulations.

While algorithmic advancements are crucial, hardware-specific optimization also plays a vital role in boosting NNPS computational efficiency. Given that NNPS algorithms search neighbors for each SPH particle individually, it is very suitable for fine-grained parallel computation on GPUs. Furthermore, low-precision computation [20-22] is progressively utilized in GPU platforms, enhancing efficiency and reducing memory demand. This is especially the case in applications where precision is not paramount, such as the early training stage of deep learning [23-25]. The specific task of searching neighbors in SPH compares distances against a critical value while distance calculation and comparison do not rely on high precision computation, making it amenable to low-precision computation for efficiency purpose. While tons of explorations and kernels have been made for examining the feasibility of using low-precision in closest neighbor searching, such as molecular dynamics simulations [26, 27] and k-nearest neighbors KNN [28-30], few comprehensive study has been conducted to the NNPS for SPH methods. Different from the other closest neighbor searching algorithms, the SPH method generally needs to search two or three layers of neighbors, *i.e.,* the searching radius of SPH is generally 2~3 folds of the particles spacing. This makes the NNPS in SPH distinct from the existing studies searching closest neighbors only.

Accordingly, this study delves into enhancing NNPS efficiency for SPH from the hardware aspect. The great advantages of GPUs over CPUs in computational efficiency have been widely demonstrated in SPH applications [31-32], where high-precision floating point 64 (FP64) is applied in the entire computation process. In this study, we examine the feasibility of integrating GPUs' low-precision computation into NNPS to optimize SPH computational efficiency. This yields a mixed-precision SPH framework that combines low-precision computation for NNPS with high-precision computation for accuracy-sensitive tasks. To assess the impact of hardware and low-precision computation, we comprehensively evaluate the performance of CPUs and GPUs across floating-point 64 (FP64), FP32, and FP16 precisions. To ensure accuracy while employing low-precision computation for efficiency purposes, we propose a relative coordinate-based cell link-list (RCLL) algorithm. Its effectiveness is examined, in this study, in 2-D scenarios, which have higher demands for significant digits compared to 3-D cases when using the same number of particles in simulations. Ultimately, this research offers insights into the interplay between hardware and algorithmic precision in the context of SPH simulations.

## SPH method

The SPH method discretizes the material domain into a set of particles, each representing a small volume and containing its own information of field variables like mass, density, velocity, location, pressure, and stress etc. SPH allows particles to move and adapt to changing flow conditions under the rigid control of SPH's governing equations.

The governing partial differential equations PDEs in SPH are typically derived from the fundamental principles of fluid mechanics, including the conservation of mass, momentum, and energy. The derived PDEs are called 'Navier-Stoke' equations in computational fluid dynamics (CFD). Yet, different from the traditional CFD methods that solves the Navier-Stoke equations under Eulerian description, the SPH methods solves the Navier-Stoke equations in Lagrangian description, which frame can naturally track the strain and deformation of material, making SPH suitable for handling solid mechanics problems that especially concern the deformation and strain-stress correlation of material.

The SPH governing equations have the form of

$$\begin{cases} \frac{D\rho}{Dt} = -\rho \frac{\partial v^\beta}{\partial x^\beta} \\ \frac{Dv^\alpha}{Dt} = \frac{1}{\rho} \frac{\partial \sigma^{\alpha\beta}}{\partial x^\beta} + f^\alpha \\ \frac{De}{Dt} = \frac{\sigma^{\alpha\beta}}{\rho} \frac{\partial v^\alpha}{\partial x^\beta} \\ \frac{Dx^\alpha}{Dt} = v^\alpha \end{cases} \quad (1)$$

where $t$ is time, $\rho$ is scalar density, $v^\alpha$ is velocity component, $e$ is internal energy, $x^\alpha$ is displacement component, $\sigma^{\alpha\beta}$ is total stress tensor, and $f^\alpha$ is body force component.

SPH method works through translating the PDEs into solvable SPH formulations by approximating the gradient of field variables using the information of nearby particles and the concept of 'kernel function' $W$:

$$\frac{\partial f(x_i)}{\partial x^\alpha} = \sum_{j=1}^{N_i} V_j f(x_j) \frac{\partial W(r_{ij}, h)}{\partial x^\alpha} \quad (2)$$

where $V_i$ is the representative volume of the target particle $i$, $N_i$ is the total number of neighboring particles within the searching radius of particle $i$, $r_{ij}$ is the distance between the target particle $i$ and its supporting particle $j$ within the searching radius. Namely, the gradient of field variable of particle $i$ can be approximated with the field variable of its neighboring particles and the gradient of kernel function $W$.

The kernel function $W$ evaluates the weight of a certain neighboring particle's contribution to the gradient approximation of the target particle. This study takes the most widely used piecewise B-spline function [11] for gradient approximation:

$$W(R, h) = \alpha_d \times \begin{cases} \frac{2}{3} - R^2 + \frac{1}{2}R^3, & 0 \leq R < 1 \\ \frac{1}{6}(2 - R)^3, & 1 \leq R < 2 \\ 0, & R \geq 2 \end{cases} \quad (3)$$

where the factor $\alpha_d = \frac{1}{h}, \frac{15}{7\pi h^2}, \frac{3}{2\pi h^3}$ in one-, two-, and three-dimensional space, respectively. $R = \frac{r}{h}$ with $r$ being the distance between two particles. $h$ is called 'smoothing length' in SPH and its value controls the searching radius ($2h$) as shown in Figure 1. Basically, $h = 1.2\Delta s$ with $\Delta s$ being the particle spacing.

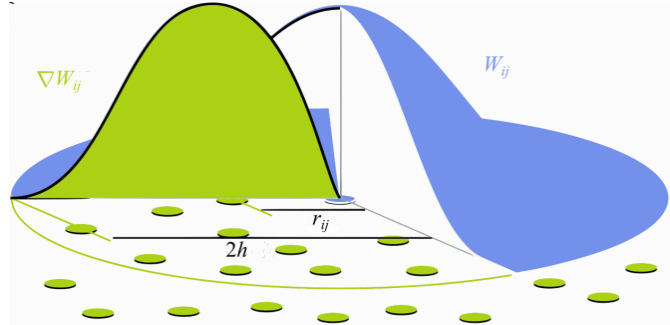

Figure 1 Illustration of kernel function in SPH method.

Conventionally, SPH expresses the volume $V_i$ in Eq. (2) as the ratio of mass and density $\frac{m}{\rho}$, yielding a discretized form of governing equations [11, 12] as

$$\begin{cases} \frac{D\rho_i}{Dt} = \sum_{j=1}^{N_i} m_j (v_i^\beta - v_j^\beta) \frac{\partial W_{ij}}{\partial x_j^\beta} \\ \frac{Dv_i^\alpha}{Dt} = \sum_{j=1}^{N_i} m_j \left(\frac{\sigma_i^{\alpha\beta}}{\rho_i^2} + \frac{\sigma_j^{\alpha\beta}}{\rho_j^2}\right) \frac{\partial W_{ij}}{\partial x_j^\beta} + f^\alpha \\ \frac{De_i}{Dt} = \frac{1}{2} \sum_{j=1}^{N_i} m_j \left(\frac{p_i}{\rho_i^2} + \frac{p_j}{\rho_j^2}\right)(v_i^\beta - v_j^\beta) \frac{\partial W_{ij}}{\partial x_j^\beta} + \frac{1}{\rho_i} \tau_i^{\alpha\beta} \varepsilon_i^{\alpha\beta} \\ \frac{Dx_i^\alpha}{Dt} = v_i^\alpha \end{cases} \quad (4)$$

where $W_{ij} = W(R_{ij}, h)$, and $p$ is pressure. SPH assumes a constant mass $m$ represented by each moving particle, meaning no mass exchange among particles.

In practical SPH simulation, some special numerical treatments must be taken in the governing PDEs to ensure the stability of SPH algorithm, such as the artificial viscosity term for removing the unphysical oscillation of variable fields [33-35] and the artificial stress term for resolving the inherent 'tensile instability' issue [36-39] of SPH.

Figure 2 plots the computation time of SPH in terms of the particle number for a case study in a Poiseuille flow simulation [40]. It clearly shows that the computational time exponentially increases with the particle number. Furthermore, the NNPS step takes a majority of computational time in SPH, especially for a large number of particles. Thus, enhancing NNPS efficiency is significant for improving SPH's efficiency by means of computation hardware as well as algorithms.

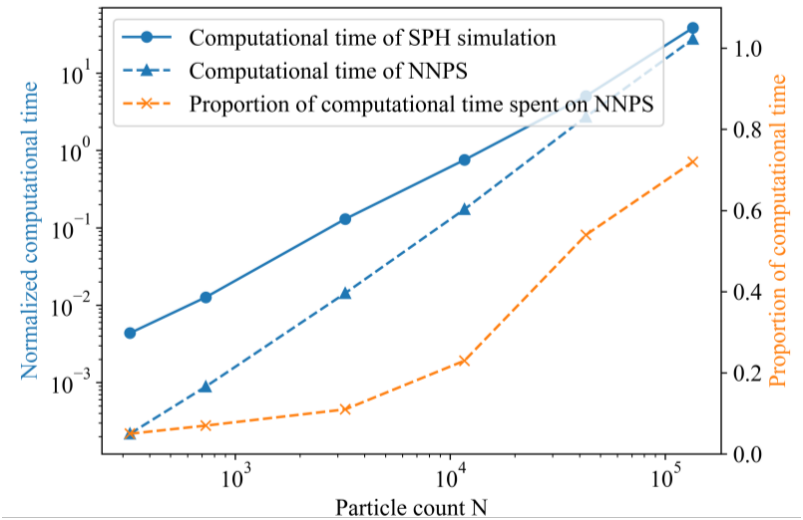

Figure 2 Proportion of computational time spent on NNPS in SPH simulation with all-list algorithm.

In this study, we select two representative NNPS algorithms: the all-list algorithm and cell link-list algorithms.

The "all-list" algorithm is a straightforward method for searching for nearest neighboring particles within a certain radius, as illustrated in Figure 3(a). In this algorithm, each particle is checked against all the other particles in the dataset to determine whether they are within the specified radius $2h$. With "$N$" particles, the total comparisons needed are $N * (N - 1)$, resulting in quadratic computation complexity O($N^2$). As particle count increases, the number of comparisons quickly escalates, making it inefficient for large numbers of particles.

In contrast, the cell link-list NNPS algorithm adopts a more efficient strategy. It starts by dividing the simulation space into a grid of uniform spatial cells, tailored to match the SPH's searching radius, as shown in Figure 3(b). Subsequently, particles are assigned to the background cells based on their spatial coordinates. This cell-assignment, performed at the simulation's onset and updated as particles move, ensures that only particles within the same cell or adjacent cells are the to-be-check particles. This approach significantly reduces the particles requiring examination, thereby enhancing efficiency and scalability. This cell-based approach leads to a computational complexity of O(*N*) because the particles to be checked in adjacent cells remain relatively fixed, irrelevant to the total particle count *N*. However, in practice, the algorithm's actual complexity can also be influenced by memory utilization. This link-list NNPS algorithm demands extra memory to maintain particle lists within the background cells, potentially increasing memory complexity. This factor becomes crucial when dealing with a substantial number (*e.g.,* over 1 million) of particles and necessitates careful consideration of memory utilization.

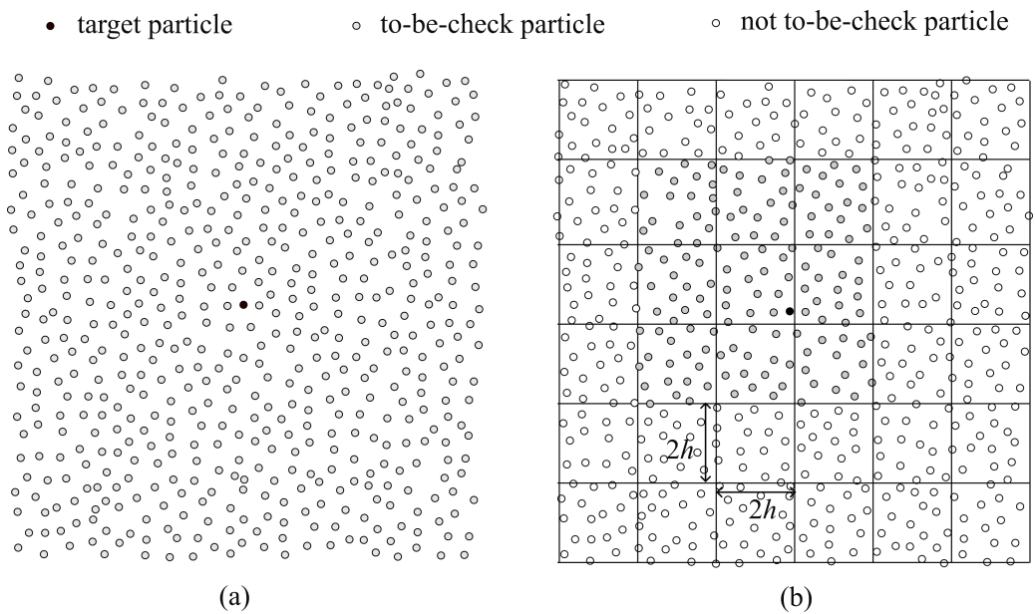

Figure 3 (a) All-list NNPS algorithm. (b) Cell link-list NNPS algorithm. In (b), the cell size equals to the searching radius $2h$ of SPH particles.

## Mixed-precision SPH

### *Low-precision NNPS algorithm*

GPU (Graphics Processing Unit) is well-suited for performing neighbor searching in SPH due to its massive parallel processing threads and high memory bandwidth. Additionally, the GPU computation efficiency of NNPS can be further enhanced by using the lower precision formats, such as the half-precision floating point (FP16) introduced in version 7.5 of CUDA [754-2008 - IEEE Standard for Floating-Point Arithmetic]. Compare to the 64-bit double-precision FP64 allowing for approximately 15 to 17 significant digits and the 32-bit single-precision FP32 providing around 7 significant digits, the FP16 employs only 16 bits that offers 3 to 4 significant digits, as shown in Figure 4, making it memory-efficient and workload-light. Theoretically, FP16 can be 2x faster than FP32, and 4x faster than FP64. In some applications with heavy memory demand, the efficiency performance of using lower precision can be even better. While lower precision can offer significant benefits in efficiency, it's crucial to carefully analyze and optimize the code to ensure the accuracy of results based on FP16 computation.

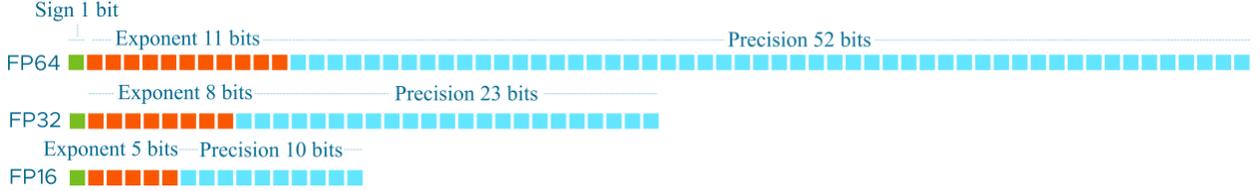

Figure 4 Comparison of FP64, FP32, and FP16 in bit length.

When applying low-precision FP16 to NNPS in SPH, accuracy can be an issue because the 3-4 significant digits provided by FP16 may not effectively store particles' coordinates requiring longer significant digits as the number of particles employed in simulations is large. This can lead to inaccuracies in NNPS results, and thereby negatively affects SPH's accuracy.

To maintain accuracy while using FP16, we propose a "relative coordinate-based link-list" or RCLL algorithm. This approach utilizes the particle-cell information of the cell link list NNPS algorithm and expresses particle coordinates in terms of cell center's coordinates and particles' relative coordinates within cells. By combining the precision of relative coordinates and cell center coordinates, we can achieve longer significant digits than using absolute coordinates alone.

For optimal use of the limited precision in FP16, both the relative coordinates of particles and the absolute coordinates of cell-centers are normalized within the range of $[-1, 1]$. Assuming the whole computational domain spans $[x_{min}, x_{max}]$ in $x$-direction, $[y_{min}, y_{max}]$ in $y$-direction, and $[z_{min}, z_{max}]$ in $z$-direction, the maximum domain length $h_d$ is defined as the maximum of these spans, *i.e.*, $h_d = \max(x_{max} - x_{min}, y_{max} - y_{min}, z_{max} - z_{min})$. Normalization is achieved through transforming original coordinate system $(x_0, y_0, z_0)$ to normalized coordinate system $(x', y', z')$ using the formula:

$$x' = \frac{2x_0 - (x_{max} + x_{min})}{h_d}, y' = \frac{2y_0 - (y_{max} + y_{min})}{h_d}, z' = \frac{2z_0 - (z_{max} + z_{min})}{h_d} \quad (5)$$

Similarly, the relative coordinates of particles within a cell can also be normalized within $[-1, 1]$ using:

$$x = 2\frac{x' - x'_{cc}}{h_c}, y = 2\frac{y' - y'_{cc}}{h_c}, z = 2\frac{z' - z'_{cc}}{h_c} \quad (6)$$

where $h_c$ represents the cell size, and $x'_{cc}, y'_{cc}, z'_{cc}$ denote the normalized absolute coordinates of cell center normalized by domain size as done in Eq. (5).

Once the normalized relative coordinates $(x, y, z)$ of particles and their cell indices are obtained, the distance of two particles $i$ and $j$ can be calculated as

$$\begin{cases} dx = (x_i - x_j)h_c + x_{cc}^I - x_{cc}^J \\ dy = (y_i - y_j)h_c + y_{cc}^I - y_{cc}^J \\ dz = (z_i - z_j)h_c + z_{cc}^I - z_{cc}^J \\ r_{ij} = \sqrt{dx^2 + dy^2 + dz^2} \end{cases} \quad (7)$$

where $(x_i, y_i, z_i)$ and $(x_j, y_j, z_j)$ are the normalized relative coordinates of particles $i$ and $j$, respectively, and $I$ and $J$ are the cell indices of particles $i$ and $j$, respectively. The whole process of RCLL is illustrated in Figure 5.

In this study, our objective is to explore the potential efficiency improvements achievable through the implementation of half-precision computation specifically in the computationally intensive neighbor-searching step. While Alonso's previous work [42] has effectively demonstrated the efficacy of such relative coordinate-based cell link list algorithm using single-precision computation, our focus is to investigate the performance of RCLL under even lower half-precision computation and evaluate the associated efficiency gains in NNPS.

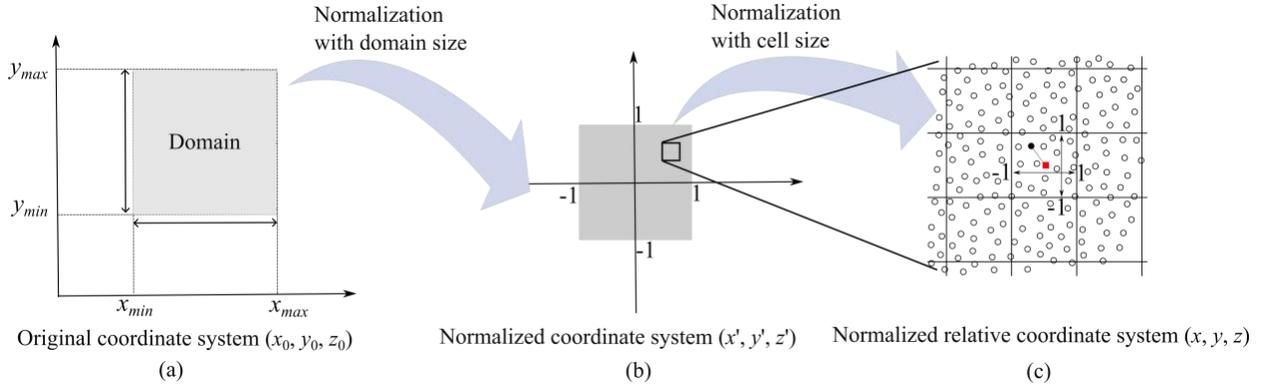

Figure 5 Illustration of normalized relative coordinates for RCLL. From (a) to (b), the original coordinates are normalized to range [-1, 1] by Eq. (5). From (b) to (c), the normalized coordinates are converted to relative coordinate in cell within range [-1,1] by Eq. (6). The red square dot in (c) denotes a cell center.

*Mixed-precision SPH framework*

The mixed-precision SPH framework is designed to utilize low-precision coordinates exclusively within NNPS while retaining high-precision coordinates for other components. However, this approach requires frequent conversions of coordinates, when converting from high precision to low precision for NNPS.

Particularly, the transitioning from high precision coordinates to low precision coordinates for NNPS needs to perform coordinate transformation represented by Eqs. (5) and (6), which process could negate the efficiency gains achieved by employing low-precision NNPS.

To address this challenge and preserve the efficiency benefits of FP16 in NNPS, our proposed mixed-precision SPH framework maintains the cell-based relative coordinates throughout the simulation. In specific, once the normalized relative coordinates are obtained at the initial time step, at subsequential steps, they can be updated directly with

$$\begin{cases} x_i(t+1) = x_i(t) + \frac{2\Delta x_i(t)}{h_c} \\ y_i(t+1) = y_i(t) + \frac{2\Delta y_i(t)}{h_c} \\ z_i(t+1) = z_i(t) + \frac{2\Delta z_i(t)}{h_c} \end{cases} \quad (8)$$

where $\Delta x$, $\Delta y$, and $\Delta z$ are the displacement components accumulated within the time duration $\Delta t$. They are calculated from velocity:

$$\Delta x_i(t) = v_i^x \Delta t, \ \Delta y_i(t) = v_i^y \Delta t, \ \Delta z_i(t) = v_i^z \Delta t \quad (9)$$

with $v_i^x, v_i^y, v_i^z$ being the three components of velocity in $x, y$, and $z$ directions, respectively, of particle $i$.

Note that, the updated relative coordinates may exceed the range of $[-1,1]$, meaning that the particle moves into an adjacent cell. In this case, the index of that adjacent cell is assigned to the particle, ensuring that relative coordinates stay within the valid range of $[-1,1]$. In this way, the normalized relative coordinates are maintained accurately, and cell indices are updated timely without the necessity of frequent coordinate transformation.

Figure 6 shows the computation flowchart of the mixed-precision SPH framework.

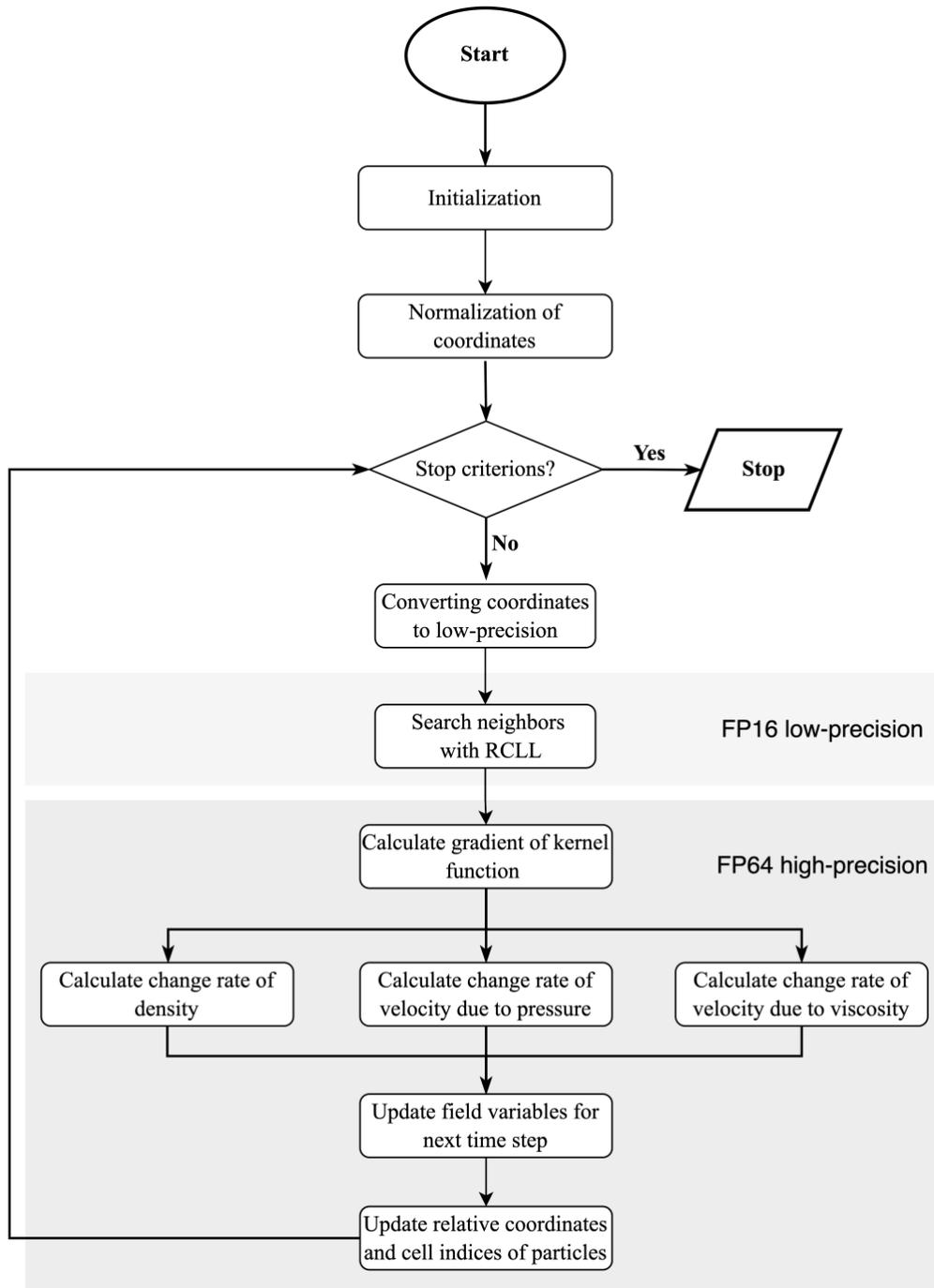

Figure 6 Flowchart of mixed-precision SPH framework.

# Results and discussions

***Efficiency of CPU vs GPU***:

In this work, we mainly use a single node of NERSC Perlmutter for the evaluation. Therefore, all the CPU works were performed on a single-core AMD EPYC 7763, while GPU works were performed on a single-core NVIDIA A100. The CUDA version 12 is used in computation.

Figure 7(a) compares CPU and GPU computational times of the all-list NNPS algorithm. It shows that GPUs yields a speedup of up to three orders of magnitude in comparison to CPUs, particularly for a large number of particles. The results in Figure 7(a) emphasizes the impact of floating-point precision on efficiency: FP32 precision enhances efficiency about two-fold compared to FP64, a trend seen on both CPU and GPU platforms. Figure 7(a) also reveals the computational complexity of the all-list algorithm, aligning with theoretical expectations of $O(N^2)$ complexity on both CPU and GPU. Yet, on GPU, small particle counts can lead to computation time worse than theoretical predictions due to memory and communication overhead. This highlights the need for GPU memory utilization optimization.

Considering the superior performance of GPU for the parallel processing NNPS algorithms, Figure 7(b) compares the GPU efficiency performances of all-list and link-list algorithms. The link-list algorithm shows a distinct advantage of $O(N)$ complexity well-suited for larger particle counts ($> 10^4$). However, for smaller particle counts ($N < 10^4$), the all-list algorithm excels due to its cell-free nature and memory efficiency. Switching from FP64 to FP32 precision brings a 2x efficiency boost for both algorithms.

While Figure 7 establishes the GPU's profound efficiency advantage over CPU, Figure 8 (a) and (b) further investigate the impact of threads count on parallel processing efficiency for all-list and link-list algorithms, respectively. Increased thread utilization reduces computational time for both, but beyond a certain threshold, adding more threads yields no extra gain. The critical thread-number threshold is higher for the more computationally intensive all-list algorithm.

Overall, the above numerical results highlight the important role of algorithm and hardware in optimizing NNPS efficiency.

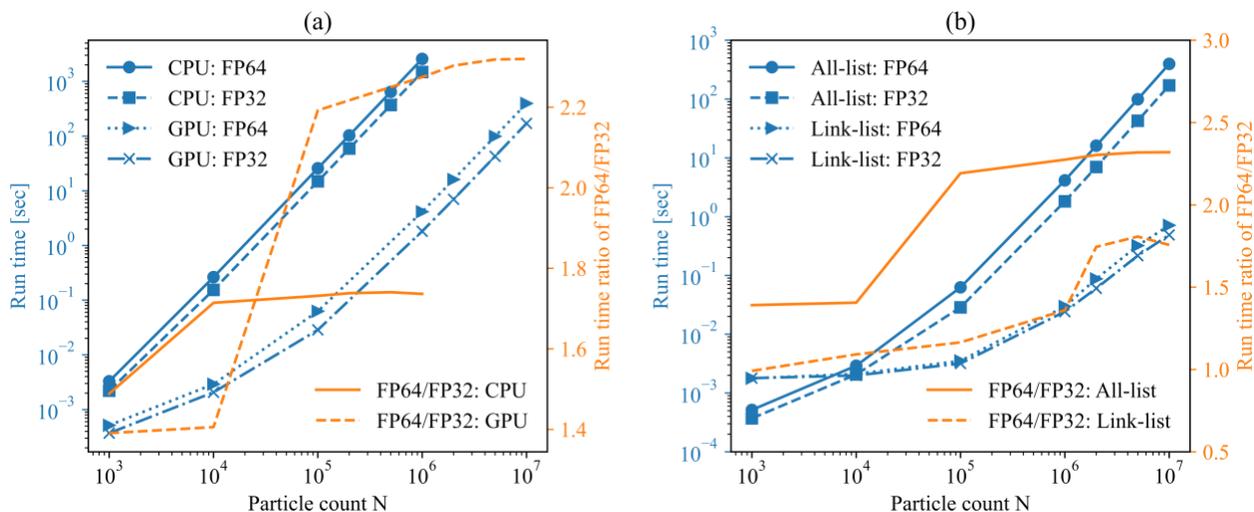

Figure 7 (a) Run time comparison of all-list algorithm on CPU *vs* GPU; (b) Run time comparison of all-list *vs* link-list algorithms on GPU.

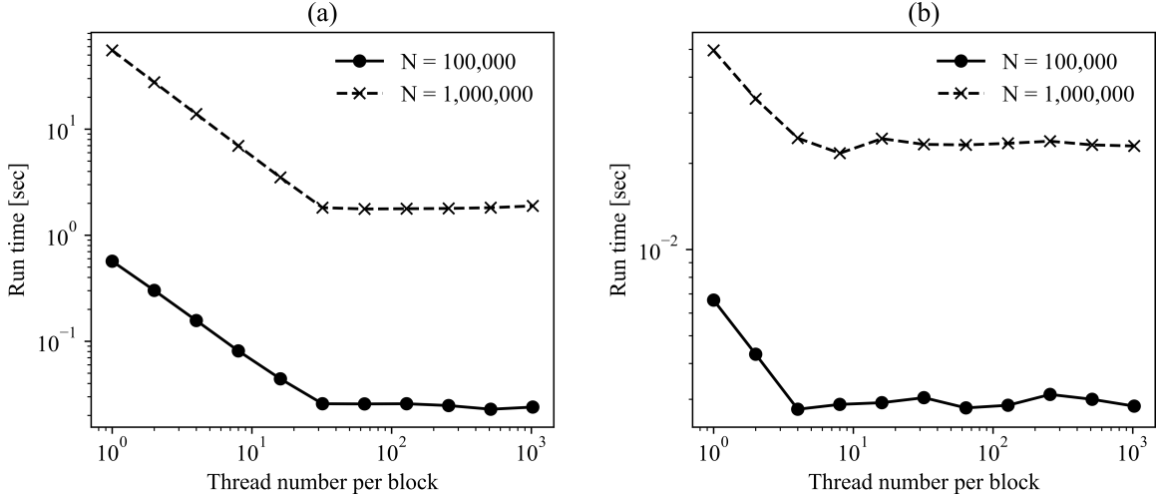

Figure 8 Effect of threads number on GPU efficiency. (a) All-list algorithm; (b) Link-list algorithm.

*Effect of half-precision on accuracy*

Given the limited significant digits of FP16 precision, a rigorous accuracy examination is essential before delving into potential efficiency gains. We meticulously assess the impact of FP16 precision on the accuracy of *neighbor searching*, *SPH gradient approximations*, and *simulations of dynamic problems*.

*Effect of FP16 on neighbors searching accuracy.* Figure 9 illustrates particle distribution configurations used to assess the accuracy of neighbor determinations using FP16 precision for distance calculations. Particles around a circular position (radius $R = 1$) with disturbances of $\Delta R$ were analyzed. Results in Table 1 indicate that while FP32 provides accurate neighbor predictions for $\Delta R > 10^{-7}$, FP16 accuracy is maintained only for $\Delta R \geq 10^{-3}$. This aligns with FP16's limited significant digits (3~4 compared to FP32's 7~8). Situations needing more decimal digits than FP16 could result in unreliable outcomes, highlighting FP16's inaccuracy for particle spacings below $10^{-3}$ within a computational domain.

To substantiate this implication, we search neighbors for all particles within a unity square based on FP16 computation. Particles randomly distribute within the unity domain, yielding an effective particle spacing $\Delta s = \sqrt[d]{1/N}$, where $d$ represents dimensions of the problem and $N$ indicates the particle count. The results listed in Table 2 demonstrate the inaccuracy of FP16 computation when $\Delta s < 10^{-3}$. In other words, FP16 can be problematic when applied to scenarios with over $10^6$ particles in 2D cases or over $10^9$ particles in 3D scenarios. While the link-list algorithm equates to the all-list algorithm in terms of neighbor determinations, the Relative Coordinate-based Link-List (RCLL) algorithm, based on relative coordinates and normalizations, outperforms, given its more decimal digits.

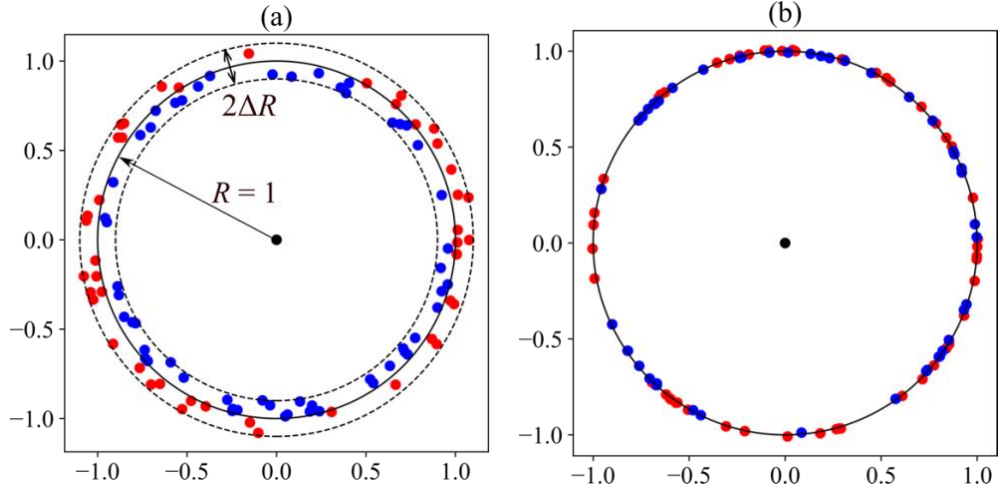

Figure 9 Distribution of particles for evaluating the accuracy of FP16 computation in 2D. (a) $\Delta R = 0.1$; (b) $\Delta R = 0.01$.

Table 1 Count of incorrect neighbor determinations by NNPS algorithm with FP16 computation.

| ΔR / Precision | $10^{-1}$ | $10^{-2}$ | $10^{-3}$ | $10^{-4}$ | $10^{-6}$ | $10^{-8}$ |
|---|---|---|---|---|---|---|
| **FP32** | 0 | 0 | 0 | 0 | 0 | 39 |
| **FP16** | 0 | 0 | 1 | 34 | 53 | 57 |

Table 2 Percentage of incorrect neighbor determinations by NNPS algorithms with FP16.

| Δs / NNPS | 0.01 | 0.005 | 0.002 | 0.00125 | 0.001 | 0.0005 |
|---|---|---|---|---|---|---|
| **All-list** | 0 | 0.028 | 0.17 | 1.2 | 3.3 | 31 |
| **Link-list** | 0 | 0.028 | 0.17 | 1.2 | 3.3 | 31 |
| **RCLL** | 0 | 0 | 0 | 0 | 0 | 0 |

*Effect of FP16 on gradient approximation accuracy*. Numerical results in Figure 10 and Table 3 reveal that FP16 precision in NNPS does not affect the 1st order accuracy of SPH gradient approximation using the normalized gradient operator in Appendix. Even at $\Delta s = 0.0005$, the error of gradient approximation with the FP16-generated inaccurate neighbors list *via* the all-list and link-list algorithm only slightly deviates and maintains the 1st order accuracy. This robustness can be attributed to two factors. First, the inaccurately detected neighbors always locate around the boundary of the supporting domain of SPH particle, with minimal contribution to the gradient approximation. Secondly, the normalized gradient operator in Appendix ensures the 1st order accuracy independent of the neighbor's selection. Nevertheless, precise detection of supporting particles is still essential for accurately simulating the interaction of SPH particles and ensuring numerical stability.

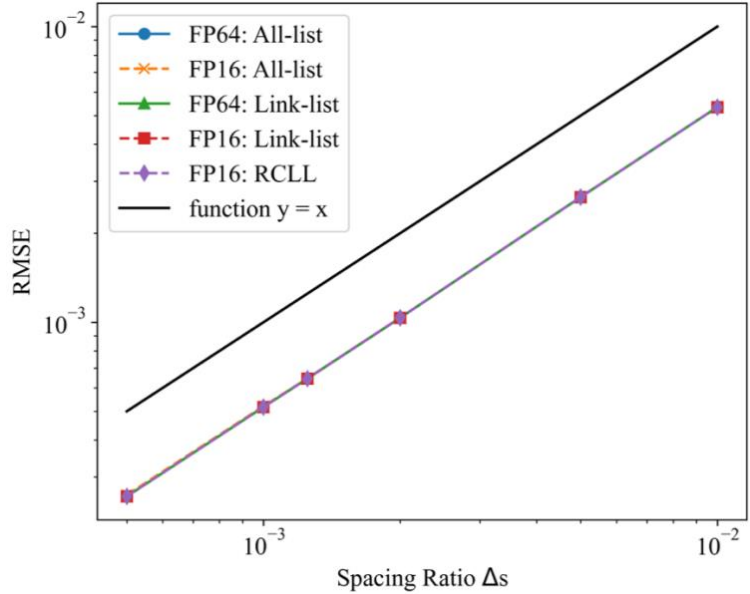

Figure 10 Accuracy comparison of high-precision FP64 *vs* low-precision FP16 computations in NNPS. The root mean square error (RMSE) is calculated as $RMSE = \sqrt{\sum_{i=1}^{N}(actual_i - predicted_i)^2 / N}$.

Table 3 Numerical error of NNPS algorithms with high-precision and low-precision computations for gradient approximation of $f(x) = x^3$ using SPH.

| Δs<br>Precision | 0.01 | 0.005 | 0.002 | 0.00125 | 0.001 | 0.0005 |
|---|---|---|---|---|---|---|
| **FP64: all-list** | 5.34E-3 | 2.65E-3 | 1.035E-3 | 6.45E-4 | 5.17E-4 | 2.58E-4 |
| **FP16: all-list** | 5.34E-3 | 2.65E-3 | 1.036E-3 | **6.46E-4** | **5.19E-4** | **2.61E-4** |
| **FP64: link-list** | 5.34E-3 | 2.65E-3 | 1.035E-3 | 6.45E-4 | 5.17E-4 | 2.58E-4 |
| **FP16: link-list** | 5.34E-3 | 2.65E-3 | 1.036E-3 | **6.46E-4** | **5.19E-4** | **2.61E-4** |
| **FP16: RCLL** | 5.34E-3 | 2.65E-3 | 1.035E-3 | 6.45E-4 | 5.17E-4 | 2.58E-4 |

*Effect of FP16 on accuracy of SPH simulation.* While FP16 precision exhibits minimal influence on gradient calculations despite notable errors in neighbor determination, its impact on the accuracy of dynamic SPH solutions needs to be examined carefully. We choose the 2D Poiseuille flow as an example, considering its simplicity and theoretical benchmarking capability [42], and take the same parameters and configurations as the previous work [40].

We compare one high-precision approaches (I) against two mixed-precision approaches (II and III), as outlined in Table 4. Approach II employs FP16 absolute coordinates in NNPS, while approach III adopts the proposed FP16 normalized relative coordinates. The accuracy of SPH solutions is evaluated by measuring velocity field and particles' location discrepancy from analytical solution [40].

Numerical tests in Figure 11 and Figure 12 demonstrate that FP16 maintains SPH solution accuracy remarkably well for particle spacings up to $\Delta s = 0.025$, corresponding to 40 particles per unit length, 1,600 particles in a unit 2D domain, or 64,000 particles in a unit 3D domain. Namely, FP16 works well for preserving accuracy across the spectrum of testing approaches for low-resolution cases. However, in high-resolution cases characterized by smaller particle spacings ($\Delta s$) and larger particle counts, using FP16 in computations could potentially affect SPH solution accuracy. Table 5 shows that the approach II, utilizing low-precision FP16 for NNPS, yields considerable discrepancy of particles' movement, especially for high-resolution cases ($\Delta s < 0.001$).

In contrast, the approach III, utilizing FP16 normalized relative coordinates in NNPS, exhibits desirable accuracy, mirroring exactly the high-precision reference solution by approach I. This is attributed to the FP16-based RCLL algorithm's capability of generating totally accurate neighbor determinations. Overall, this signifies RCLL's robustness in upholding SPH solution accuracy, paving the way for the utilization of low-precision FP16 to achieve efficiency goals while maintaining accuracy.

Table 4 Configurations of three numerical tests for examining the influence of FP16 on the accuracy of SPH solutions to the 2D Poiseuille flow.

| Approach / Section | NNPS algorithm | Float precision in NNPS | Float precision in other computations |
|---|---|---|---|
| **I** | Link-list | FP64 | FP64 |
| **II** | Link-list | FP16 | FP64 |
| **III** | RCLL | FP16 | FP64 |

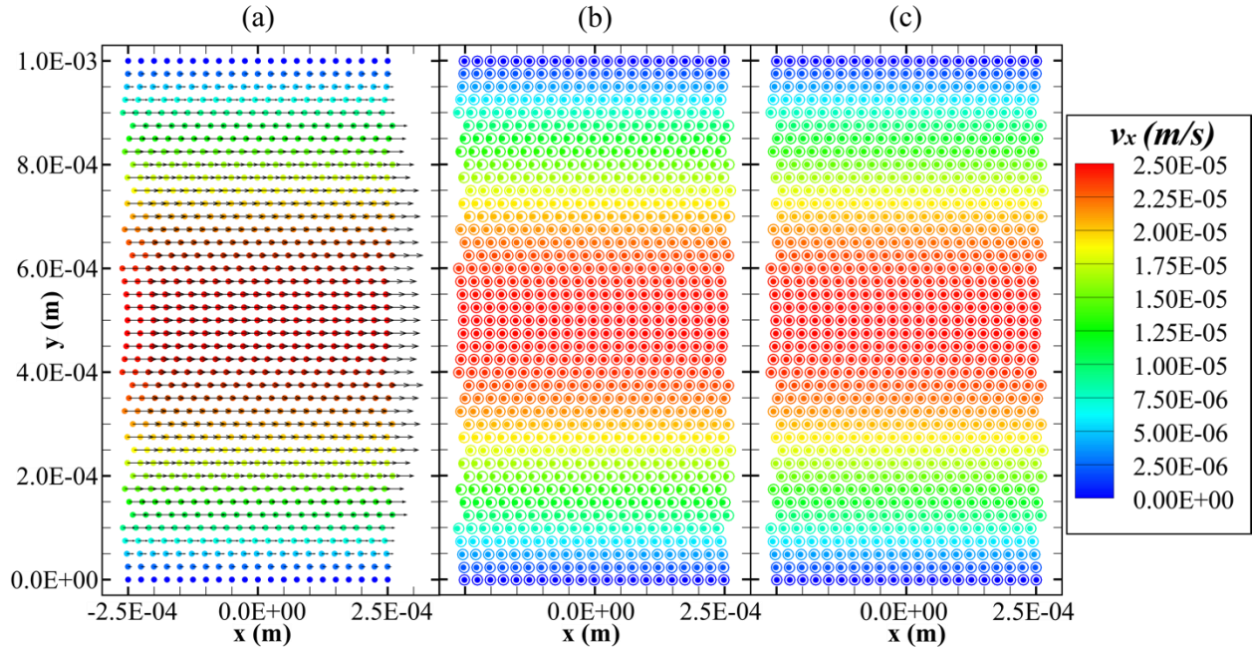

Figure 11 Contour of velocity field from SPH simulations of the Poiseuille flow. (a) Velocity vector and velocity contour from SPH simulations with link-list algorithm by FP64 computation; (b) Comparison of particles distribution from SPH simulations with approach I (null circles) and apprach II (solid dots); (c) Comparison of particles distribution from SPH simulations with approach I (null circles) and approach III (solid dots).

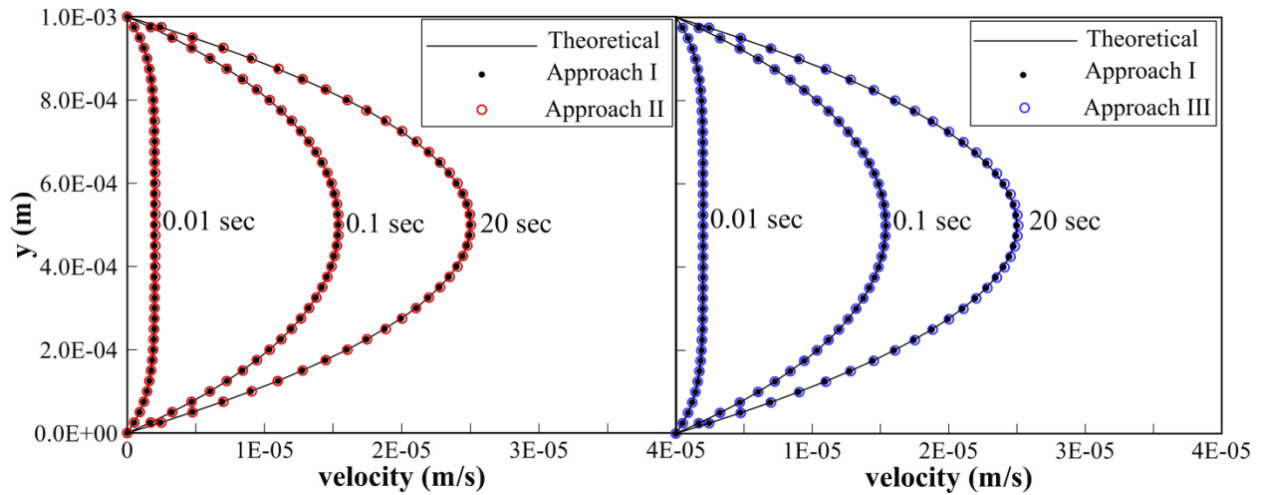

Figure 12 Comparison of velocity profiles from SPH simulations with approach II (a) and approach III (b) to the theoretical solution and the SPH solution with approach I.

Table 5 Maximum location discrepancy of particles in the whole domain at $t = 1$ sec by the three approaches compared to the theoretical solution . The theoretical solution of accumulated displace is computed by integrating the theoretical velocity solution [40] over time.

| Δs<br>Approach | $0.025$ | $0.01$ | $0.0025$ | $0.001$ | $0.00025$ | $0.0001$ |
|---|---|---|---|---|---|---|
| I | 0.11Δs | 0.13Δs | 0.16Δs | 0.13Δs | 0.14Δs | 0.16Δs |
| II | 0.15Δs | 0.22Δs | 0.27Δs | 1.3Δs | 3.4Δs | 9.5Δs |
| III | 0.11Δs | 0.13Δs | 0.16Δs | 0.13Δs | 0.14Δs | 0.16Δs |

*Effect of half-precision on efficiency*

Figure 13 and Figure 14 illustrate how floating-point precisions affect the computational efficiency of the all-list and RCLL algorithms, respectively. From Figure 13(a), we observe that FP32 is 2x faster than FP16 and FP16 is further 2x faster than FP32 in NNPS as expected. Using FP16 can achieve an overall ~5x efficiency improvement in SPH, as shown in Figure 13(b). This efficiency gain is mainly attributed to the dominant computational time spent on the NNPS process when employing the computationally intensive $O(N^2)$ all-list NNPS algorithm. In contrast, Figure 14 shows that using FP16 in the RCLL algorithm yields only ~1.5x efficiency enhancement, much lower than the theoretical expectation. This discrepancy arises because the computationally lighter $O(N)$ RCLL algorithm does not dominates the SPH computation as done by the all-list algorithm, and overhead from other sources plays a more substantial role in computation.

Figure 15 compares the RCLL algorithm's efficiency across different precisions in 3-D cases. The findings indicate that employing half-precision FP16 in RCLL for 3-D scenarios brings an approximate 1.7x enhancement in efficiency over the standard double-precision FP64. This efficiency enhancement in 3-D cases is superior to that in 2-D cases as presented in Figure 14, primarily attributed to the increased proportion of computational time spent on NNPS in 3D cases. In 3-D scenarios, a larger number of particles are subjected to neighbor-checking than 2-D cases. Nevertheless, the efficiency enhancement achieved by using low-precision FP16 is still much lower than the theoretical value.

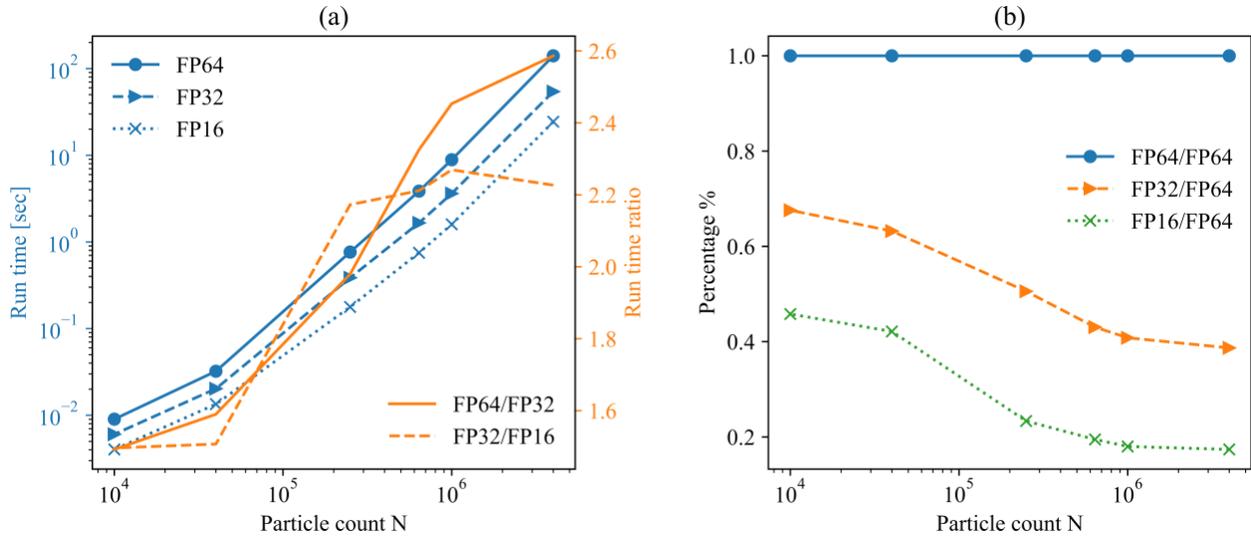

Figure 13 (a) Run time comparison of all-list NNPS algorithm under FP64 *vs* FP32 *vs* FP16. (b) Run-time ratio of SPH using low FP precisions in all-list NNPS algorithm to that using high precision FP64.

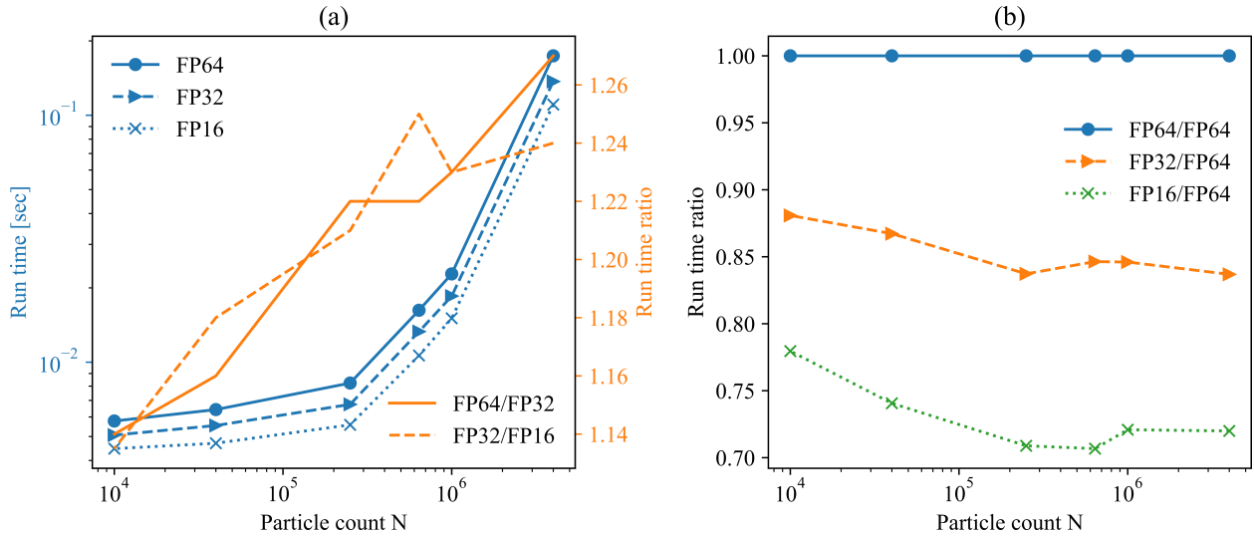

Figure 14 (a) Run time comparison of RCLL algorithm under FP64 *vs* FP32 *vs* FP16. (b) Run-time ratio of SPH using low float precisions in RCLL algorithm to that using FP64.

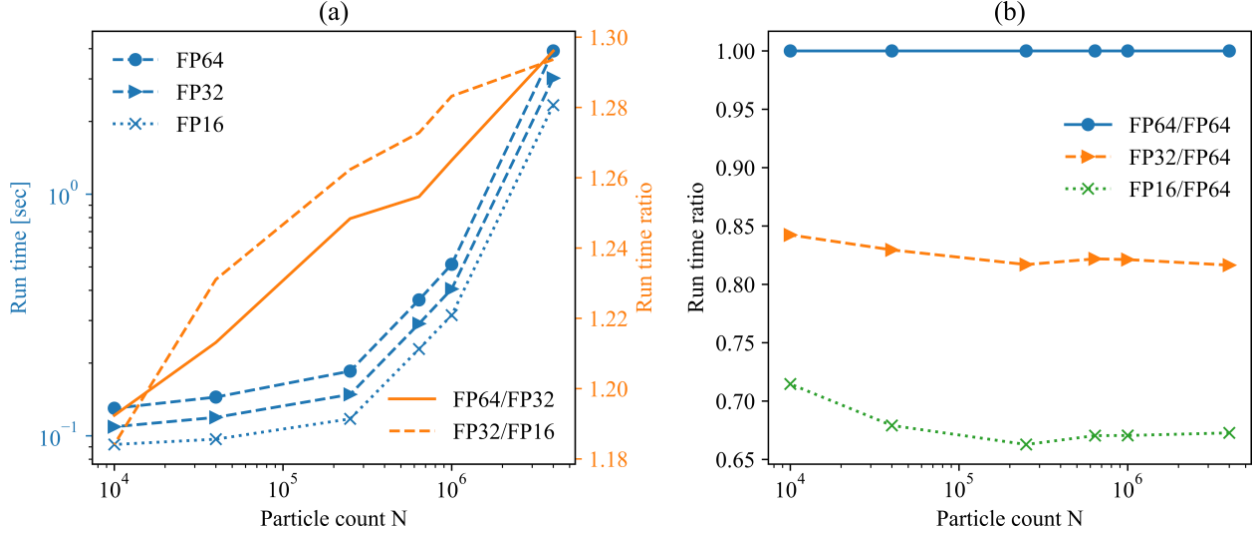

Figure 15 (a) Run time comparison of RCLL algorithm under FP64 *vs* FP32 *vs* FP16 in 3D examples. (b) Run-time ratio of SPH using low float precisions in RCLL algorithm to that using FP64 in 3D examples.

To identify the primary factor limiting the performance of the FP16 RCLL algorithm, we utilized the official NVIDIA Nsight profiler to collect performance metrics of our CUDA program, as listed in Table 6. The data indicates that the RCLL-based NNPS process consumes only 51% of the available memory bandwidth and utilizes a mere 8% of the computational capacity, which suggests the NNPS kernel is primarily memory-bound. Therefore, optimizing memory usage becomes a critical target for performance improvement.

Upon close examination, we found that the arbitrary ordering of particles makes the concurrently running GPU threads handle spatially distant particles, resulting in varying to-be-check neighboring cells among threads and thus inefficient memory loads. The data locality significantly influences the data accessing efficiency in SPH computation [43]. To enhance memory utilization and boost parallelism, we take advantage of NVIDIA GPUs' capability to cache global memory accesses in L1/L2, which feature enables more efficient memory access when repeatedly accessing the same memory region within a single execution context. Our optimization strategy involves sorting particles based on their spatial coordinates and having concurrently running threads within a warp search neighbors for those particles that are close to each other and have the same to-be-check particles. This treatment greatly increases the chances of data reuse from the cache and reduces the memory accessing region, thus greatly enhancing computational efficiency. In implementation, we first sorted the particles according to their *x* coordinates, and then performed a secondary sort based on their *y* coordinates for particles with identical *x* values through the CUDA Thrust library.

This optimization significantly improves the compute throughput of NNPS from 8.31 to 30.15, as listed in Table 6, yielding a remarkable 2.7x efficiency improvement in NNPS. Figure 16 clearly illustrates the enhanced memory utilization achieved through this optimization. Thus, it confirms the significance of memory management as a potential bottleneck in the performance of computationally efficient NNPS algorithms, suggesting more attention and efforts required in the algorithms aspect to fully use GPU memory capacity.

Table 6 Profiling information of CUDA code evaluated by Nsight. The evaluated program calculates the gradient of a cubic function with the FP16 RCLL-based SPH method where 1 million particles are employed in computation. In specific, the 'NNPS' kernel searches neighbors while the 'gradient approximation' kernel calculates gradient using the SPH gradient operator based on the identified neighbors list by the NNPS kernel.

| Kernel name | Duration (millisecond) before/after optimization | Compute throughput(%) before/after optimization | Memory throughput(%) before/after optimization |
|---|---|---|---|
| NNPS | 7.0/2.60 | 8.31/30.2 | 51.3/59.8 |
| Gradient approximation | 1.79/1.32 | 20.0/51.5 | 90.9/57.6 |

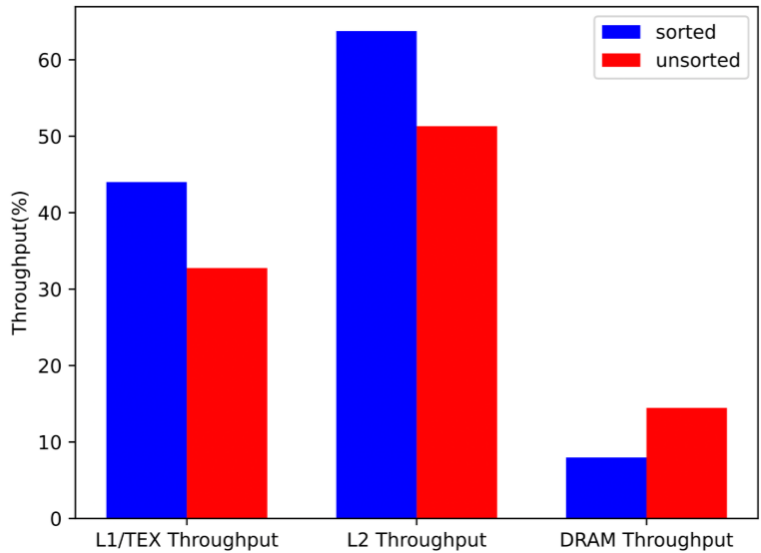

Figure 16 Comparison of Cache and global memory throughput before optimization (unsorted) and after optimization (sorted).

At last, we examined the computational efficiency performance of the FP16-based RCLL algorithm on the AMD MI250, which is a counterpart to the NVIDIA A100 GPU. The testing results, as shown in Figure 17, indicates that the two GPUs exhibit very similar runtimes when executing the algorithm. When comparing FP16 computation to FP32 computation, it is worth noting that FP16 offers a slightly higher efficiency enhancement on the NVIDIA platform than on the AMD platform. In essence, both AMD and NVIDIA GPUs demonstrate comparable efficiency performance within the context of our mixed-precision SPH framework.

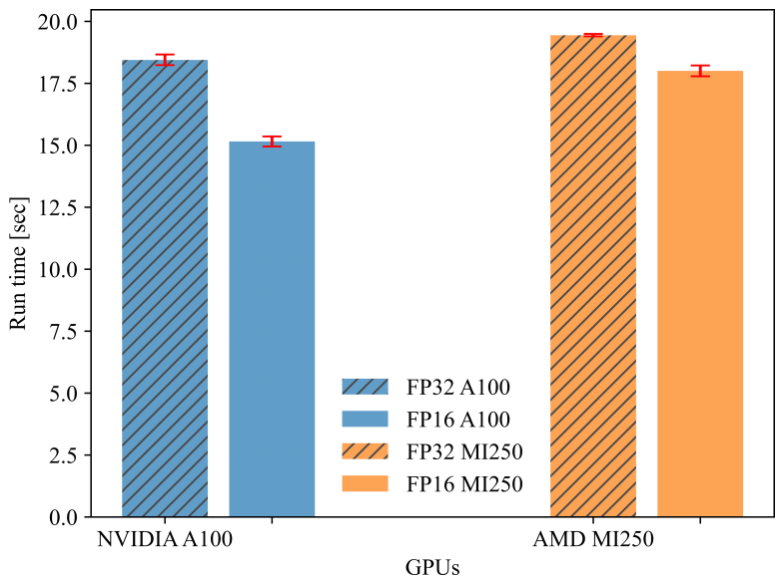

Figure 17 Comparison of RCLL algorithm's run times on NVIDIA A100 vs AMD MI250 when 1 million particles are employed in simulation.

## Conclusions and discussions

In summary, our investigation digs into the potential efficiency gains we can achieve from the most updated hardware-specific advancements and optimizations, including CPUs vs. GPUs, the role of float-point precision, and the memory management.

The Nearest Neighboring Particle Searching (NNPS) algorithm is proven well-suited for GPU parallelization, where GPUs exhibit a remarkable three-order-of-magnitude speedup over CPUs in NNPS tasks. The strategic utilization of lower float-point (FP) precision in NNPS substantially enhances computational efficiency. Yet, it is crucial to exercise caution when using ultra-low precision, such as FP16, as it may cause inaccuracy issue, particularly in high-resolution situations. Our Relative Coordinated-based Link List (RCLL) NNPS algorithm emerges as a desirable solution, capable of maintaining commendable accuracy even under FP16 precision.

The impact of reduced precision is most sensitive in the computationally demanding all-list NNPS algorithm, characterized by an $O(N^2)$ complexity. Here, FP16 precision yields a remarkable fourfold efficiency enhancement. Conversely, the more efficient $O(N)$ RCLL algorithm demonstrates a milder 1.5x efficiency boost, mainly due to the influence of overhead from other sources, especially the memory utilization. Advanced memory optimization strategies, including a proper spatial sorting of particles and efficient data caching, propel a further 2.7x efficiency improvement for RCLL algorithm.

Notwithstanding these advancements, the intricacies associated with managing a variable number of particles in each cell pose challenges for optimizing shared memory utilization. This variability necessitates a dynamic memory allocation strategy, which can introduce complexities in ensuring optimal usage of available memory resources. Effectively addressing this issue remains a paramount concern, as it represents a key bottleneck in fully utilizing the potential efficiency gains offered by GPUs. Moreover, algorithm-level enhancements, such as reusing previously computed distances, hold promise for further efficiency enhancement. By strategically storing and retrieving previously calculated distance data, redundant calculations can be minimized, reducing the computational workload. However, it's essential to recognize that implementing such techniques necessitates meticulous memory management, as efficiently storing and accessing this cached data within a GPU's memory hierarchy can be intricate. Balancing the benefits of reduced computation against the overhead of memory management complexities is a critical consideration in pursuing this avenue of optimization.

In conclusion, efficient SPH simulations require a careful balance between hardware, algorithms, and precision. Memory optimization strategies hold promise for further improving GPU-based SPH simulations.

# Appendix

Applying the Tylor expansion of function $f$ at particle $j$ in terms of the information of nearby particle $i$ to the standard SPH gradient operator in Eq. (2) yields,

$$\langle\frac{\partial f_i}{\partial x}\rangle = \sum_{j=1}^{N_i} V_j \left( f_i + f_i^x \Delta x + \frac{f_i^{xx}}{2!} \Delta x^2 + \cdots \right) \frac{\partial W_{ij}}{\partial x} \tag{A1}$$

Since the $i$-related terms are independent of the summation index $j$, they can be extracted out of the summation operation in Eq. (A1), giving

$$\langle\frac{\partial f_i}{\partial x}\rangle = f_i \sum_{j=1}^{N_i} V_j \frac{\partial W_{ij}}{\partial x} + f_i^x \Delta x \sum_{j=1}^{N_i} V_j \frac{\partial W_{ij}}{\partial x} + O(\Delta x^2) \tag{A2}$$

Based on a simple algebraic work, we can easily get the gradient $f_i^x$ on particle $i$ with

$$f_i^x = \frac{\langle\frac{\partial f_i}{\partial x}\rangle - f_i \sum_{j=1}^{N_i} V_j \frac{\partial W_{ij}}{\partial x}}{\Delta x \sum_{j=1}^{N_i} V_j \frac{\partial W_{ij}}{\partial x}} + O(\Delta x) \tag{A3}$$

Considering the formulation of $\langle\frac{\partial f_i}{\partial x}\rangle$ in Eq. (2) and the relation $\Delta x = x_j - x_i$, we get

$$f_i^x = \frac{\sum_{j=1}^{N_i} V_j(f_j-f_i)\frac{\partial W_{ij}}{\partial x}}{\sum_{j=1}^{N_i} V_j(x_j-x_i)\frac{\partial W_{ij}}{\partial x}} + O(\Delta x) \tag{A4}$$

In this formulation, the C0 consistency of gradient approximation has no special requirement on the volume term, which can simply be considered as a constant.

Thus, we can get a 1st-order accurate volume-free SPH gradient operator:

$$\begin{cases} \langle f_i^x \rangle = \frac{\sum_{j=1}^{N_i}(f_j-f_i)\frac{\partial W_{ij}}{\partial x}}{\sum_{j=1}^{N_i}(x_j-x_i)\frac{\partial W_{ij}}{\partial x^\alpha}} \\ \langle f_i^y \rangle = \frac{\sum_{j=1}^{N_i}(f_j-f_i)\frac{\partial W_{ij}}{\partial y}}{\sum_{j=1}^{N_i}(y_j-y_i)\frac{\partial W_{ij}}{\partial y}} \\ \langle f_i^z \rangle = \frac{\sum_{j=1}^{N_i}(f_j-f_i)\frac{\partial W_{ij}}{\partial z}}{\sum_{j=1}^{N_i}(z_j-z_i)\frac{\partial W_{ij}}{\partial z}} \end{cases} \tag{A5}$$

## Author Contributions:

Conceptualization: Ang Li, Zirui Mao, Ganesh Gopalakrishnan; methodology: Zirui Mao; investigation: Zirui Mao, Xinyi Li; supervision: Ang Li, Ganesh Gopalakrishnan; discussion: Zirui Mao, Xinyi Li, Ang Li, Ganesh Gopalakrishnan, Shenyang Hu; writing – original draft preparation: Zirui Mao, Xinyi Li; writing – review & edit: Ang Li, Shenyang Hu, Ganesh Gopalakrishnan; funding acquisition: Ang Li, Ganesh Gopalakrishnan.

## Acknowledgements

This material is based upon work supported by the U.S. Department of Energy, Office of Science, Office of Advanced Scientific Computing Research, ComPort: Rigorous Testing Methods to Safeguard Software Porting, under Award Number 78284. This research used resources of the National Energy Research Scientific Computing Center (NERSC), a U.S. Department of Energy Office of Science User Facility located at Lawrence Berkeley National Laboratory, operated under Contract No. DE-AC02-05CH11231. This research used resources of the Oak Ridge Leadership Computing Facility, which is a DOE Office of Science User Facility supported under Contract DE-AC05-00OR22725. The Pacific Northwest National Laboratory is operated by Battelle for the U.S. Department of Energy under Contract DE-AC05-76RL01830.